\documentclass[a4paper,11pt]{article}
\pdfoutput=1
\usepackage{jcappub}
\usepackage[utf8]{inputenc}
\usepackage[T1]{fontenc}
\usepackage{microtype}
\usepackage{bm}
\usepackage{soul}
\usepackage[dvipsnames,svgnames,x11names]{xcolor}
\usepackage{aas_macros}
\usepackage{multirow}
\usepackage{physics}
\usepackage{mathrsfs}

\usepackage{tensor}

\usepackage[normalem]{ulem}

\usepackage[export]{adjustbox}

\providecommand{\sorthelp}[1]{}

\begin{document}

\title{Quadratic shape biases in three-dimensional halo intrinsic alignments}

\newcommand{\IPMU}{Kavli Institute for the Physics and Mathematics of the Universe (WPI), 
UTIAS, The University of Tokyo, Kashiwa, Chiba 277-8583, Japan}
\newcommand{\IAS}{School of Natural Sciences, Institute for Advanced Study, 1 Einstein Drive, Princeton, NJ 08540, USA}
\newcommand{\ASIAA}{Academia Sinica Institute of Astronomy and Astrophysics (ASIAA), No. 1, Section 4, Roosevelt Road, Taipei 10617, Taiwan}

\author[a]{Kazuyuki Akitsu,}
\author[b]{Yin Li}
\author[c,d]{and Teppei Okumura}

\affiliation[a]{\IAS}
\affiliation[b]{Department of Mathematics and Theory, Peng Cheng
Laboratory, Shenzhen, Guangdong 518066, China}
\affiliation[c]{\ASIAA}
\affiliation[d]{\IPMU}

\abstract{
Understanding the nonlinear relation between the shapes of halos or galaxies and the surrounding matter distribution is essential in accurate modeling of their intrinsic alignments.
In the perturbative treatment, such nonlinear relation of the intrinsic alignments appears as higher-order shape bias parameters.
In this paper, we present accurate measurements of the quadratic shape bias parameters by combining the \emph{full three-dimensional} power spectrum of the intrinsic alignments (i.e., without any projection) with the quadratic field method.
In order to benefit from the full three-dimensional power spectrum we employ the spherical tensor decomposition of the three-dimensional shape field and measure their power spectra for the first time.
In particular, we detect the vector and tensor power spectra in this basis, which cannot be explained by the widely-used nonlinear alignment model.
Further, by cross-correlating the three-dimensional halo shape field with the quadratic shape bias operators from the initial condition of the same simulation to cancel cosmic variance, we effectively extract bispectrum information and detect quadratic shape bias parameters in the intrinsic alignments with high significance for the first time.
We also compare these measurements with the prediction where quadratic shape biases are dynamically generated from the linear Lagrangian shape bias through the large-scale bulk flow.
We find general agreement for all three biases with small deviations, which in practice could be negligible for the current photometric surveys.
This implies that the advection prediction for the higher-order shape biases can be used as a prior in the cosmological analyses of intrinsic alignments.
}

\date{\today}

\maketitle

\section{Introduction}

The large-scale structure of the Universe has been a cornerstone of cosmology for decades. 
The distribution of matter and galaxies on the largest scales provides crucial insights into the fundamental properties of the Universe, such as its geometry, composition, and evolution. 
Over time, small fluctuations in the matter density generated by the cosmic inflation grew under the influence of gravity to form the vast cosmic web of filaments, voids, and clusters, which we observe today. 
As a result, the distribution of galaxies exhibits correlations that are explained by the combination of the initial conditions and gravitational evolution. 
Similarly, recent studies also show that galaxy (or halo) shapes are correlated with their surrounding large-scale structure, known as the intrinsic alignments (IA)~\cite{Catelan:2000vm,Hirata:2004gc,Mandelbaum:2005hr,Hirata:2007np,Okumura:2008bm,Okumura:2008du,Singh:2014kla}.

While the intrinsic alignments used to be considered as a contaminant to the weak lensing~\cite{Catelan:2000vm,Hirata:2004gc,Bridle:2007ft,Joachimi:2010xb}, it has recently been realized that the intrinsic alignments itself contains a wealth of cosmological information~\cite{Chisari:2013dda, Schmidt:2012nw, Schmidt:2013gwa, Schmidt:2015xka, Akitsu:2020jvx, Taruya:2020tdi,Kogai:2020vzz, Akitsu:2022lkl, Okumura:2021xgc,Okumura:2023pxv, Kurita:2023qku}.
In both cases, we need to properly model this phenomenon in order to extract unbiased cosmological information.
A major challenge here is to understand various non-linearities as in the galaxy clustering case.
Fortunately, we can recycle the basic framework developed in the context of galaxy clustering to study these non-linearities, since the underlying physics that governs the intrinsic alignments should be similar.
In particular, the perturbative approach to the intrinsic alignments attracts attention given its rigorousness and the success in the galaxy clustering~\cite{Blazek:2011xq, Blazek:2015lfa,Vlah:2019byq, Vlah:2020ovg, Matsubara:2022eui, Matsubara:2022ohx, Bakx:2023mld, Matsubara:2023avg}.

In this paper, we focus on the nonlinear biasing among several non-linearities.
The most commonly used model of IA is the linear alignment \cite{Catelan:2000vm} where the shape of galaxy or halo is assumed to be linearly aligned with the large-scale tidal field, corresponding to the linear bias description in terms of the bias expansion.
However, the extension to include the higher-order shape biases would be mandatory to obtain an accurate model for IA, given the nonlinear nature of structure formation~\cite{Blazek:2011xq,Blazek:2015lfa}.
In fact, Ref.~\cite{Schmitz:2018rfw} pointed out that higher-order shape biases are inevitably introduced due to the advection, as discussed in the galaxy clustering bias~\cite{Baldauf:2012hs,Chan:2012jj,Desjacques:2016bnm}.
These higher-order biases are important ingredients in modeling the bispectrum and one-loop power spectrum of IA.
In particular, the tree-level bispectrum requires the quadratic shape biases and one-loop power spectrum requires quadratic and cubic biases.
Although these biases are necessary components to complete spectra beyond the linear power spectrum, varying these biases induces degeneracies between parameters and generally degrades the parameter constrains~\cite{Bakx:2023mld}.
Hence, it would be important to obtain prior knowledge of higher-order shape biases both from the theoretical grounding and from comparison with cosmological simulations.
The goal of this paper is to get constraints on the quadratic shape bias parameters from $N$-body simulations and to compare the measurements with the theoretical prediction from the advection argument, discussed in Ref.~\cite{Schmitz:2018rfw}.

There are several ways to measure bias parameters from simulations.
The most common method is to compare halo-matter (or galaxy-matter) spectra with matter spectra on large-scales\footnote{There is another way to accurately measure renormalized bias parameters using the separate universe simulations. See Refs.~\cite{Li:2015jsz,Baldauf:2015vio,Lazeyras:2015lgp} for the density bias measurements and Refs.~\cite{Stucker:2020fhk,Akitsu:2020fpg} for the linear shape bias measurements.}.
For instance, the linear density bias can be estimated from the halo-matter power spectrum in large-scale limit, and the quadratic density bias from the halo-matter-matter bispectrum in large-scale limit~\cite{Saito:2014qha}.
Similarly, the linear shape bias can be estimated from the large-scale shape-matter power spectrum~\cite{Kurita:2020hap}, and the quadratic shape bias parameters would be estimated from the large-scale shape-matter-matter bispectrum.
However, since measuring the bispectrum is computationally expensive we take another route to estimate the quadratic shape bias parameters, which is the quadratic field method proposed in Ref.~\cite{Schmittfull:2014tca} and used to estimate the cubic density bias parameters in Ref.~\cite{Lazeyras:2017hxw, Abidi:2018eyd}.

The quadratic field method fully utilizes the fact that we know the initial conditions of the simulation.
Because in the $N$-body simulation we have the initial conditions at the field level, we can construct a basis of quadratic bias operators at the field level.
Cross-correlations between the basis from the initial conditions and the late-time halo density field can be expressed as a set of bias parameters and a particular combination of cross-correlations of quadratic operators with themselves, which are also measurable from the same initial conditions, leading to the cosmic variance cancellation.

We generalize this technique to be applicable to the halo shape field while taking full advantage of $N$-body simulations where we can measure three-dimensional halo shape.
Although previous studies use the projected shape because it is the actual observable, the IA itself happens in three-dimensional space.
Therefore it is better to use full three-dimensional information to calibrate IA models.
To this end we use the spherical tensor decomposition of a rank-2 tensor (corresponding to the 3D shape field) \cite{Vlah:2019byq,Vlah:2020ovg,Matsubara:2022ohx,Matsubara:2022eui,Matsubara:2023avg}.
We measure the power spectra of the halo shape field in spherical tensor basis for the first time, and further combine this decomposition with the quadratic field method to constrain the quadratic shape bias parameters.

The rest of the paper is organized as follows: in Sec.~\ref{sec:method}, we briefly discuss the bias expansion for the shape field, including how the higher-order shape biases are expected to appear, introduce the spherical tensor basis to efficiently obtain full three-dimensional shape information, and illustrate the quadratic field method extended to the tensor field.
We then show the measurements of the power spectra of the spherical tensor fields and the quadratic shape biases in Sec.~\ref{sec:result}.
Finally we conclude in Sec.~\ref{sec:discussion}.

\section{Method}
\label{sec:method}

In this section we first provide a brief review on the bias expansion of the shape field and how the higher-order biases in Eulerian space arise from the linear Lagrangian bias via advection.
Then, we introduce the spherical tensor decomposition of a rank-2 tensor to efficiently capture full three-dimensional information of the shape field.
Finally we discuss the quadratic field method to measure the quadratic shape biases in simulations.

\subsection{Shape bias expansion}
\label{subsec:shape_bias}

In general, we can expand the galaxy or halo shape in fields constructed from the matter density field as\footnote{${S_{ij}^{\rm (E)}}$ is generally not traceless as we will briefly discuss in Sec.~\ref{subsec:simulations} and the bias expansion of the trace fluctuation follows the usual density bias expansion with different bias values. In this paper, we focus on the trace-free part of this rank-2 tensor, which corresponds to the usual shape field and hence $S_{ij}$ should be interpreted as the trace-free tensor, unless otherwise stated.}
\cite{Schmitz:2018rfw}
\begin{align}
    S^{\rm (E)}_{ij}({\bf x}) = b_K^{\rm (E)} K_{ij}({\bf x})
    + b^{\rm (E)}_{\delta K} \delta({\bf x}) K_{ij}({\bf x})
    + b^{\rm (E)}_{KK} \left[KK\right]_{ij}({\bf x})
    + b^{\rm (E)}_T T_{ij}({\bf x}),
    \label{eq:bias_expansion_Eulerian}
\end{align}
up to second order, where we have introduced
\begin{align}
    K_{ij}({\bf x}) \equiv & \left[\frac{\partial_i\partial_j}{\partial^2} - \frac13 \delta_{ij}^{\rm K}\right] \delta({\bf x}) \equiv {\cal D}_{ij}\delta({\bf x})
    \\
    \left[KK\right]_{ij}({\bf x}) \equiv & K_{ia}({\bf x})K_{aj}({\bf x}) - \frac13 \delta_{ij}^{\rm K}K^2({\bf x})
    \\
    T_{ij}({\bf x}) \equiv & {\cal D}_{ij}\left( \delta^2({\bf x}) - \frac32 K^2({\bf x}) \right) = \frac{21}{4} {\cal D}_{ij}\left(\delta^{(2)}({\bf x}) - \theta^{(2)}({\bf x})  \right),
\end{align}
with $K^2({\bf x}) \equiv K_{ij}({\bf x})K^{ij}({\bf x})$ and $\delta^{(2)}$ and $\theta^{(2)}$ being the second-order density and velocity divergence in the standard perturbation theory~\cite{Bernardeau:2001qr, Desjacques:2016bnm}.
The first term in Eq.~\eqref{eq:bias_expansion_Eulerian} is the only term that appears in the linear alignment model~\cite{Catelan:2000vm, Hirata:2004gc}.
The superscript $^{\rm (E)}$ emphasizes that we write down the bias expansion for the \textit{Eulerian} shape field.
We expect to have the quadratic \textit{Eulerian} shape biases even when there are no second-order \textit{Lagrangian} shape biases due to the advection~\cite{Schmitz:2018rfw,Taruya:2021jhg}, as in the galaxy density bias case~\cite{Baldauf:2012hs,Chan:2012jj}.

Before explaining the advection effect we note the effect of the density-weighting.
Above we define the halo or galaxy shape biases as if the shape is a volume-weighted quantity. 
However, in reality we can measure it only where a galaxy or halo is.
Thus the measured shape field naturally becomes density-weighed:
\begin{align}
    \Tilde{S}^{\rm (E)}_{ij}({\bf x}) = S^{\rm (E)}_{ij}({\bf x})[1+\delta^{\rm (E)}_{\rm g}({\bf x})],
\end{align}
where 
the tilde denotes the density-weighted quantity and 
$\delta_{\rm g}({\bf x})$ is the overdensity field of a biased tracer.
This should also be the case in Lagrangian space:
\begin{align}
    \Tilde{S}^{\rm (L)}_{ij}({\bf q}) = S^{\rm (L)}_{ij}({\bf q})[1+\delta^{\rm (L)}_{\rm g}({\bf q})],
\end{align}
with ${\bf q}$ being the initial Lagrangian coordinates and $S^{\rm (L)}_{ij}({\bf q})$ and $\delta_{\rm g}^{\rm (L)}({\bf q})$ being the \textit{Lagrangian} shape and density fields, respectively.
The higher-order shape bias expansion is subject to this density-weighting: to be precise, although the basis for the bias expansion remains the same, their specific values would be different. At second order, for instance, if we define the shape biases for the density-weighted shape field, it acquires the following contribution,
\begin{align}
    \Tilde{S}^{\rm (L)}_{ij}({\bf q}) \supset b_1^{\rm (L)}b_K^{\rm (L)}\delta({\bf q}) K_{ij}({\bf q}),
    \hspace{0.5cm}{\rm or}\hspace{0.5cm}
    \Tilde{S}^{\rm (E)}_{ij}({\bf x}) \supset b_1^{\rm (E)}b_K^{\rm (E)}\delta({\bf x}) K_{ij}({\bf x})
\end{align}
where we have used $\delta_{g}= b_1 \delta$ at this order.
Thus, the shape bias of the density-weighting term becomes
\begin{align}
    \Tilde{b}_{\delta K} = b_{\delta K} + b_1 b_K.
    \label{eq:tilde_bdK}
\end{align}
We will come back to this issue in Sec.~\ref{subsec:quad_shape_bias}, where we show the measurements of the shape biases.

Our starting point to discuss the advection is the relation,\footnote{This is how we define the Lagrangian shape field. In fact, it is not trivial how the Lagrangian (initial) galaxy/halo shape should transform to the Eulerian (final) shape field or vice versa. For instance, the Lagrangian tidal field, $K^{\rm (L)}_{ij}({\bf q}) \sim (\partial_{{\bf q}_i}\partial_{{\bf q}_j}/\partial^2_{\bf q})\delta^{\rm (L)}({\bf q})$, is not related to the Eulerian tidal field, $K^{\rm (E)}_{ij}({\bf x}) \sim (\partial_{{\bf x}_i}\partial_{{\bf x}_j}/\partial^2_{\bf x})\delta^{\rm (E)}({\bf x})$, in this way (Eq.~\eqref{eq:Lagrangian_Eulerian_shape}). The assumption here is that the Lagrangian shape is aligned with the Eulerian shape field.}
\begin{align}
    \Tilde{S}^{\rm (E)}_{ij}({\bf x}) = \int \dd^3 {\bf q}~ \delta_{\rm D}^{(3)}({\bf x-q-\Psi({q})}) 
    ~ \Tilde{S}_{ij}^{\rm (L)}({\bf q}),
    \label{eq:Lagrangian_Eulerian_shape}
\end{align}
which is equivalent to write
\begin{align}
    \Tilde{S}^{\rm (E)}_{ij}({\bf x}) ~\dd^3 {\bf x} = \Tilde{S}^{\rm (L)}_{ij}({\bf q}) ~\dd^3 {\bf q},
    \label{eq:shape_mapping}
\end{align}
where the displacement field, $\Psi({\bf q})$, maps a biased tracer from the Lagrangian coordinates ${\bf q}$ to the Eulerian coordinates ${\bf x}$.
As the same mapping holds for the density, we also have 
\begin{align}
    \left[1+\delta_{\rm g}^{\rm (E)}({\bf x})\right] \dd^3 {\bf x} = 
    \left[1+\delta_{\rm g}^{\rm (L)}({\bf q})\right] \dd^3 {\bf q}.
    \label{eq:density_mapping}
\end{align}
Using Eq.~\eqref{eq:density_mapping} to eliminate the Jacobian in Eq.~\eqref{eq:shape_mapping} we obtain
\begin{align}
    S_{ij}^{\rm (E)}({\bf x}) = S_{ij}^{\rm (L)}({\bf q}).
\end{align}
Expanding the right-hand side up to the second order yields
\begin{align}
    S_{ij}^{\rm (E)}({\bf x}) =&~ S_{ij}^{\rm (L)}({\bf x}) - {\bf \Psi}\cdot\nabla S^{\rm (L)}_{ij}({\bf x})
    \nonumber\\
    =&~ S_{ij}^{\rm (L)} ({\bf x}) - b_K^{\rm (L)} {\bf \Psi}\cdot\nabla K_{ij}({\bf x})
    \nonumber\\
    =&~ b_K^{\rm (L)}\left[ K_{ij}^{(1)}({\bf x}) + K_{ij}^{(2)}({\bf x}) \right]
    \label{eq:2nd_expansion_Euler}\\
    &~+\left[b_{\delta K}^{\rm (L)} - \frac23 b_K^{\rm (L)}\right]\delta({\bf x})K_{ij}({\bf x})
    +\left[b_{KK}^{\rm (L)} - b_K^{\rm (L)} \right]\left[KK\right]_{ij}({\bf x})
    +\left[b_T^{\rm (L)} -\frac{10}{21}b_K^{\rm (L)} \right]T_{ij}({\bf x}),
    \nonumber
\end{align}
where in the last equality we have used (see e.g. \cite{Desjacques:2016bnm,Schmitz:2018rfw})
\begin{align}
    K^{(2)}_{ij} = {\cal D}_{ij} \delta^{(2)} =  \frac{10}{21}T_{ij}
    + \left[KK\right]_{ij}
    + \frac23 \delta K_{ij}
    - {\bf \Psi}\cdot\nabla K_{ij}.
    \label{eq:K2ij_decomposition}
\end{align}
Comparing Eq.~\eqref{eq:2nd_expansion_Euler} with Eq.~\eqref{eq:bias_expansion_Eulerian} reads
\begin{align}
    b_K^{\rm (E)} = b_K^{\rm (L)},\quad b_{\delta K}^{\rm (E)} = b_{\delta K}^{\rm (L)} - \frac23b_K^{\rm (L)},\quad
    b_{KK}^{\rm (E)} = b_{KK}^{\rm (L)} - b_K^{\rm (L)},\quad b_T^{\rm (E)} = b_T^{\rm (L)} - \frac{10}{21}b_K^{\rm (L)}.
    \label{eq:2nd_bias_LtoE}
\end{align}
In contrast to the density case, the Eulerian linear shape bias is identical to the Lagrangian one because the tracelss tidal field can not induce any volume distortion at leading order.
At the absence of the quadratic shape biases in Lagrangian space, i.e., $b_{\delta K}^{\rm (L)}=b_{KK}^{\rm (L)}=b_T^{\rm (L)}=0$,
Eq.~\eqref{eq:2nd_bias_LtoE} gives the quadratic shape biases as a function of the linear shape bias:
\begin{align}
    b^{\rm (E)}_{\delta K} = -\frac23 b_K^{\rm (E)},\quad b_{KK}^{\rm (E)} = - b_K^{\rm (E)},\quad b_T^{\rm (E)} = - \frac{10}{21}b_K^{\rm (E)}.
    \label{eq:quad_shape_advection}
\end{align}
Note that these relations hold for the shape biases defined with respect to the volume-weighted shape field.
For the density-weighted shape field, we have
\begin{align}
    \Tilde{b}^{\rm (E)}_{\delta K} = b_1^{\rm (E)}b_K^{\rm (E)} - \frac23 b_K^{\rm (E)},
    \label{eq:quad_shape_advection_density_weighting}
\end{align}
and the other two relations remain unchanged.
Any deviation from these relations implies the presence of the quadratic shape bias in Lagrangian space, which we will explore in Sec.~\ref{subsec:quad_shape_bias}.

\subsection{Spherical tensor decomposition}
\label{subsec:ST_decomposition}

Here we summarize the spherical tensor decomposition of three-dimensional tensor fields.
Given a symmetric tensor field $X_{ij}$, we can decompose it as
\begin{align}
    X_{ij}({\bf k}) = \sum_{\ell=0,2}\sum_{m=-\ell}^{\ell} {\mathcal X}_{\ell m}({\bf k}) \left[Y^{\ell m}({\hat{k}})\right]_{ij},
\end{align}
where $\left[Y^{\ell m}({\hat{k}})\right]_{ij}$ is the spherical tensor basis, defined up to $\ell=2$ as
\begin{align}
    \left[Y_{\ell=0}^{m=0}({\hat{k}})\right]_{ij} 
    &\equiv \sqrt{\frac13}\delta_{ij}^{\rm K},
    \nonumber\\
    \left[Y_{\ell=2}^{m=0}({\hat{k}})\right]_{ij} 
    &\equiv \sqrt{\frac32} \left[ \hat{k}_i\hat{k}_j - \frac13 \delta_{ij}^{\rm K} \right],
    \nonumber\\
    \left[Y_{\ell=2}^{m=\pm 1}({\hat{k}})\right]_{ij} 
    &\equiv \sqrt{\frac12} \left[ \hat{k}_i e^\pm_j + \hat{k}_j e^\pm_i \right],
    \nonumber\\
    \left[Y_{\ell=2}^{m=\pm 2}({\hat{k}})\right]_{ij} 
    &\equiv e^\pm_i e^\pm_j,
\end{align}
in which
\begin{align}
    {\bm e}^\pm \equiv \frac{1}{\sqrt{2}}\left({\bm e}_1\mp i {\bm e}_2 \right),
\end{align}
with some orthonormal choice of $\{{\bm k}, {\bm e}_1, {\bm e}_2\}$.
Here ${\bm e}^{\pm}$ is the (contravariant) helicity basis satisfying ${\bm e}^{\pm}\cdot{\bm e}^{\pm} = 0$ and ${\bm e}^{\pm}\cdot{\bm e}^{\mp} = {\bm e}^{\pm}\cdot\left({\bm e}^{\pm}\right)^* = 1$.
As a result, the spherical tensor basis is also orthonormal
\begin{align}
    \left[Y^{\ell m}(\hat{k})\right]_{ij} \left[Y^{\ell' m'}(\hat{k})\right]^{*}_{ij} = \delta_{\ell\ell'}^{\rm K}\delta_{mm'}^{\rm K},
    \label{eq:ST_orthogonal}
\end{align}
so the helicity component ${\mathcal X}_{\ell m}({\bf k})$ can be obtained from $X_{ij}({\bf k})$ as
\begin{align}
    {\mathcal X}_{\ell m}({\bf k}) = \sum_{ij} X_{ij}({\bf k})\left[Y^{\ell m}({\hat{k}})\right]^*_{ij}.
\end{align}
Thanks to the rotational symmetry, we have 
\begin{align}
    \langle {\mathcal X}_{\ell m} ({\bf k}) {\mathcal Y}_{\ell' m'} ({\bf k}') \rangle = 
    (2\pi)^3 \delta^{\rm K}_{mm'} \delta_{\rm D}({\bf k}+{\bf k}') P^m_{XY;\ell\ell'}(k),
    \label{power}
\end{align}
for general tensor fields $X$ and $Y$.

Let us denote the helicity component of a three-dimensional galaxy shape $S_{ij}$ by ${\mathcal S}_{\ell m}$.
In this paper, we focus on the ellipsoidal component in the shape, i.e., we only consider the trace free part of the shape tensor.
In terms of the helicity basis, this means that only $\ell=2$ components are relevant.\footnote{The $\ell=0$ component corresponds to the galaxy size.}
At the linear order we have 
\begin{align}
    {\mathcal S}^{(1)}_{20} & = \sqrt{\frac23} b_K \delta  = b_K {\mathcal K}^{(1)}_{20},
    \\
    {\mathcal S}^{(1)}_{2\pm 1} &= {\mathcal S}^{(1)}_{2\pm 2} =  0,
\end{align}
where we have defined ${\mathcal K}^{(1)}_{\ell m}({\bf k})\equiv \sum_{ij}K^{(1)}_{ij}({\bf k})\left[Y_{\ell m}(\hat{k}) \right]_{ij}^*$.
As in the standard cosmological perturbation theory, the scalar-sourced tidal field only generates the scalar ($m=0$) mode.
While there is only an $m=0$ mode in the linear alignment model, we have $m=\pm1$ and $m=\pm2$ modes via the second order shape bias terms, as we will see below.
This is analogous to the fact that in the cosmological perturbation theory, vector and tensor modes arise from the scalar perturbations beyond the second order.

We introduce the spherical tensors, $\left[\delta{\mathcal K} \right]_{2 m}$, $\left[{\mathcal K{\mathcal K}} \right]_{2 m}$, ${\mathcal T}_{2 m}$, and $\left[ \Psi\cdot\nabla{\mathcal K}\right]_{2m} $, for the quadratic operators, $\delta K_{ij}$, $[KK]_{ij}$, $T_{ij}$, and ${\bf \Psi}\cdot\nabla K_{ij}$, respectively. 
Because $K_{ij}\propto [Y_{2}^{m=0}(\hat{k})]_{ij}$ and $T_{ij}\propto [Y_{2}^{m=0}(\hat{k})]_{ij}$ by definition,
${\cal K}_{2m}$ and ${\cal T}_{2m}$ have only the $m=0$ component thanks to the orthonomality of the spherical tensor basis (Eq.~\eqref{eq:ST_orthogonal}).
Hence, 
${\mathcal T}_{2\pm1} = {\mathcal T}_{2\pm2}=0$ and ${\mathcal K}^{(2)}_{2\pm 1}={\mathcal K}^{(2)}_{2\pm 2}=0$, while the other components do not vanish.
This implies that for $m=\pm 1$ and $m=\pm 2$ there is a constraint
\begin{align}
    \left[{\mathcal K{\mathcal K}} \right]_{2m} + \frac23 \left[\delta{\mathcal K} \right]_{2 m} - \left[ \Psi\cdot\nabla{\mathcal K}\right]_{2m} = 0,
    \label{eq:consistency_ST_basis}
\end{align}
which we use to validate our spherical tensor power spectrum estimations.
Furthermore, for $m=\pm 1$ one can show that 
\begin{align}
    \left[{\mathcal K{\mathcal K}} \right]_{2\pm 1}
    = \frac13 \left[\delta{\mathcal K} \right]_{2 \pm 1} 
    = \frac13 \left[ \Psi\cdot\nabla{\mathcal K}\right]_{2 \pm 1}.
    \label{eq:identity_pm1}
\end{align}
We provide the proof of this identity in Appendix \ref{app:proof}.

In terms of the spherical tensor basis, the shape bias expansion Eq.~\eqref{eq:bias_expansion_Eulerian} is now
\begin{align}
    {\cal S}_{2m} & = 
    b_K {\mathcal K}^{(1)}_{2m}
    +
    b_K {\mathcal K}^{(2)}_{2m}
    + b_{\delta K} \left[\delta{\mathcal K} \right]_{2 m}
    + b_{KK} \left[{\mathcal K} {\mathcal K} \right]_{2 m}
    + b_{T} {\mathcal T}_{2 m}
    \\
    &= b_K {\mathcal K}^{(1)}_{2m}
    +
    \sum_{\cal O} b_O {\cal O}_{m},
    \label{eq:Slm_bias_expansion}
\end{align}
where we have introduced $O \in \{ K^{(2)}, \delta K, KK, T \}$.
Note that the shape fields with $m=\pm1,$ or $\pm 2$ contain only quadratic fields other than $K^{(2)}_{ij}$ and $T_{ij}$, and they survive only if the quadratic shape biases (namely, $b_{\delta K}$ and $b_{KK}$) are nonzero.
Hence, the detection of correlations between the $m\neq 0$ fields would provide evidence of the quadratic shape biases.
We will consider correlations of the shape field with quadratic fields constructed from the Gaussian linear density field to optimally measure the quadratic shape biases.

\subsection{Quadratic field method}
\label{subsec:quad_field_method}

In this section we extend the quadratic field method proposed in Ref.~\cite{Schmittfull:2014tca} to (spherical) tensor fields.
At second order, while we have three free shape bias parameters, we have four independent quadratic tensor fields, namely, $\delta K_{ij}$, $[KK]_{ij}$, $T_{ij}$, and ${\bf \Psi}\cdot\nabla K_{ij}$; there is no free bias for the field representing the effect of the displacement (or advection) due to the equivalence principle.
Projecting onto the spherical tensor basis, the Fourier space expression of these fields reads
\begin{align}
    {\cal Q}^{W_R}_{m}[\delta_{\rm lin}]({\bf k})
    =\sum_{ij}\left[Y^{2 m}({\hat{k}})\right]^*_{ij}\int_{\bf q} \delta_{\rm lin}({\bf q})\delta_{\rm lin}({\bf k-q}) 
    M^{Q}_{ij}({\bf q}, {\bf k-q})
    W_R({\bf q}) W_R({\bf k-q}),
    \label{eq:spherical_tensor_quad}
\end{align}
where ${\cal Q}_{m} \in \{\left[\delta{\mathcal K} \right]_{2 m}, \left[{\mathcal K{\mathcal K}} \right]_{2 m}, {\mathcal T}_{2 m}, \left[ \Psi\cdot\nabla{\mathcal K}\right]_{2m}\}$ and $M^{Q}_{ij}$ is the symmetrized coupling kernel in Fourier space for each quadratic tensor field. 
The explicit forms of the each kernel are
\begin{align}
    M^{\delta K}_{ij}({\bf p},{\bf q}) &= 
    \frac12 \left(\frac{p_ip_j}{p^2} + \frac{q_iq_j}{q^2}\right)
    +\frac13 \delta_{ij}^{\rm K},
    \\
    M^{KK}_{ij}({\bf p},{\bf q}) &= 
    \frac12 \mu_{pq}\frac{p_iq_j + p_jq_i}{pq} 
    - \frac13 \left(\frac{p_ip_j}{p^2} + \frac{q_iq_j}{q^2}\right)
    -\frac13 \delta_{ij}^{\rm K} \left(\mu_{pq}^2 - \frac23\right),
    \\
    M^{T}_{ij}({\bf p},{\bf q}) &= 
    -\frac32 \left(\frac{k_ik_j }{k^2} -\frac13\delta_{ij}^{\rm K}\right)(\mu^2_{pq}-1),
    \\
    M^{\Psi\cdot\nabla K}_{ij}({\bf p},{\bf q}) &= 
    -\frac12 ({\bf p}\cdot{\bf q})\left[\frac{p_ip_j + q_jq_i}{p^2q^2} 
    -\frac13 \delta_{ij}^{\rm K} \left(\frac{1}{p^2}+\frac{1}{q^2}\right)\right],
\end{align}
where $\mu_{pq} = \hat{p}\cdot\hat{q}$ and ${\bf p} = {\bf k} - {\bf q}$.

Eq.~\eqref{eq:spherical_tensor_quad} shows that the quadratic fields at a given Fourier mode receive contributions from all Fourier modes, implying that the quadratic fields are affected by the small-scale modes that are not controlled by the perturbative expansion.
To avoid being affected by these small-scale modes and get the physical (renormalized) biases, we construct the quadratic fields with the smoothed initial density field.
We introduce a Gaussian smoothing in Fourier space $W_R(k) = \exp(-k^2R^2/2)$ with $R=20~{\rm Mpc}/h$, following Ref.~\cite{Abidi:2018eyd}.

Taking cross-correlations between the shape field (Eqs.~\eqref{eq:Slm_bias_expansion}) and the quadratic fields (Eq.~\eqref{eq:spherical_tensor_quad}) yields\footnote{We use the prime on the correlator to indicate that this expectation is equal to the power spectrum, i.e., $\langle X({\bf k})Y({\bf k}')\rangle' = P_{XY}(k)$ without the $2\pi$ and Dirac delta factors in \eqref{power}.}
\begin{align}
    \left\langle {\cal S}_{2 m}({\bf k}) {\cal Q}^{W_R}_{m}({\bf k}') \right\rangle^\prime
     = b_K \left\langle {\cal K}^{(1)}_{20}({\bf k}) {\cal Q}^{W_R}_{0}({\bf k}') \right\rangle^\prime + \sum_{\cal O} b_{\cal O} \left\langle {\cal O}_{m}({\bf k}) {\cal Q}^{W_R}_{m}({\bf k}') \right\rangle^\prime,
     \label{eq:cross_corre_shape_quad_fields}
\end{align}
where each correlation of the quadratic fields in the summation corresponds to the ``22'' term in the one-loop power spectrum:
\begin{align}
    \label{eq:22_PT_express}
    \left\langle {\cal O}_{m}({\bf k}) {\cal Q}^{W_R}_{m}({\bf k}') \right\rangle^\prime
     &= 2\sum_{ijkl} \left[Y^{2 m}({\hat{k}})\right]^*_{ij}\left[Y^{2 m}({\hat{k}^\prime})\right]^*_{kl}
     \\
     \nonumber
     &\hspace{0.5cm}\times \int_{\bf q} P_{\rm lin}(q)P_{\rm lin}(|{\bf k}-{\bf q}|) M^{O}_{ij}({\bf q}, {\bf k-q}) M^{Q}_{kl}({\bf q}, {\bf k-q})
    W_R({\bf q}) W_R({\bf k-q}),
\end{align}
with $M_{ij}^{O}({\bf p}, {\bf q})$ representing the coupling kernel for the $O$ fields.
One could measure the shape biases appearing on the right-hand side of Eq.~\eqref{eq:cross_corre_shape_quad_fields} by comparing the measurements of cross-power spectra between the shape field and the quadratic fields (the left-hand side) with the perturbation theory computation of Eq.~\eqref{eq:22_PT_express} (see e.g. Ref.~\cite{Bakx:2023mld} for this direction).
However, this method suffers greatly from the cosmic variance and therefore many realizations are required to beat the cosmic variance.
Instead, we obtain these cross-power spectra directly from the initial conditions of the simulations to cancel the cosmic variance.
In other words, we measure all the correlators appeared in Eq.~\eqref{eq:cross_corre_shape_quad_fields} from simulations with the same initial seeds.
In Eq.~\eqref{eq:cross_corre_shape_quad_fields}, the first term vanishes in an infinite volume limit or as ensemble average since this is an odd-correlator.
In reality, however, the finite simulation box results in the nonzero odd-correlator for a given realization and it enters as noise.
Hence, it is important to keep this term in this method to remove these contributions at the field level.

To estimate the linear shape bias $b_K$, we use the cross power spectrum of the shape field and linear tidal field projected onto $m=0$ basis, namely,
\begin{align}
    \chi^2_{\rm lin} = \sum_{k_i}^{k_{\rm max}} 
    \left( 
    \frac{\langle {\cal S}_{20}({\bf k}_i) {\cal K}^{(1)}_{20}({\bf k'}_i) \rangle^\prime/\langle {\cal K}^{(1)}_{20}({\bf k}_i) {\cal K}^{(1)}_{20}({\bf k'}_i) \rangle^\prime - b_K}{\sigma(\langle {\cal S}_{20}({\bf k}_i) {\cal K}^{(1)}_{20}({\bf k'}_i) \rangle^\prime/\langle {\cal K}^{(1)}_{20}({\bf k}_i) {\cal K}^{(1)}_{20}({\bf k'}_i) \rangle^\prime)} 
    \right)^2,
\end{align}
with $k_{\rm max} = 0.03~h/{\rm Mpc}$.
The variance in the denominator is measured from the 8 realizations of the simulations described below.
For the quadratic shape biases, we first introduce the following shorthand notation,
\begin{align}
    \Delta {\cal S}_{2m}({\bf k})  = {\cal S}_{2m}({\bf k}) - b_K {\cal K}_{20}^{(1)}({\bf k})\delta_{m0}^{\rm K} - \sum_{\cal O}b_{\cal O}{\cal O}_{m}({\bf k}),
\end{align}
to define
\begin{align}
    \chi^2_{\rm quad} = \sum_{k_i}^{k_{\rm max}}\sum_{\cal Q}\sum_{m=0}^{2}
    \left( \frac{\langle \Delta{\cal S}_{2 m}({\bf k}) {\cal Q}^{W_R}_{m}({\bf k}') \rangle^\prime}{\sigma(\langle \Delta{\cal S}_{2 m}({\bf k}) {\cal Q}^{W_R}_{m}({\bf k}') \rangle^\prime)} \right)^2,
\end{align}
with $k_{\rm max} = 0.08 ~h/{\rm Mpc}$.
The $\mathcal Q$-summation sums over all the quadratic fields, i.e., ${\cal Q}_{m} \in \{\left[\delta{\mathcal K} \right]_{2 m}, \left[{\mathcal K{\mathcal K}} \right]_{2 m}, {\mathcal T}_{2 m}, \left[ \Psi\cdot\nabla{\mathcal K}\right]_{2m}\}$.
Again, all the correlators appeared in these $\chi^2$ are measured from the simulations and their corresponding initial conditions to reduce the sample variance.
We search the best-fit bias parameters that minimize the joint likelihood, $\chi^2 = \chi^2_{\rm lin} + \chi^2_{\rm quad}$, by running MCMC.
The uncertainties of the measurements of the shape bias parameters shown in the next section are 68$\%$ confidence regions of marginalized posteriors of each bias parameters, estimated from the MCMC samples, with very wide uninformative priors.

\subsection{Simulations}
\label{subsec:simulations}

We perform $N$-body simulations in a cubic box of length $L=1.5~{\rm Gpc}/h$ with $N_{\rm p}=1536^3$ dark matter particles by running \texttt{L-Gadget2}~\cite{Springel:2005mi}  with $N_{\rm mesh} = 3072^3$ TreePM grid.
The initial positions and velocities are computed with the second-order Lagrangian perturbation theory at $z=49$ using \texttt{CLASS}~\cite{Blas:2011rf}  and \texttt{2LPTIC}~\cite{Crocce:2006ve} .
The cosmological parameters we adopt are consistent with the Planck result~\cite{Planck:2015fie}: $\Omega_{\rm m}=0.3089$, $\sigma_8 = 0.8149$, and $H_0 = 67.74$.
We generate eight independent realizations to estimate uncertainties.

Dark matter halos are identified with two different halo finder, \texttt{Rockstar}~\cite{Behroozi:2011ju} and \texttt{AHF}~\cite{Knollmann:2009pb} to study the dependence of the quadratic shape biases on the halo finder, where the former uses the phase-space friends-of-friends algorithm while the latter uses the spherical overdensity to identify dark matter halos.
We also employ several shape definitions to study the dependence of the quadratic shape biases on the shape definition.
For both the Rockstar and AHF halos, we measure each halo shape using the reduced and original inertia tensor, 
\begin{align}
    I^{\rm red}_{ij} = \frac{1}{N_{\rm p}}\sum_n \frac{x_{n;i}x_{n;j}}{x^2_n}, 
    \hspace{1cm}
    I^{\rm nonred}_{ij} = \frac{1}{N_{\rm p}}\sum_n x_{n;i}x_{n;j}, 
\end{align}
where $x_{n;i}$ is the position vector of each member dark matter particle from the halo center.
In addition, for the Rockstar halos we measure the halo shape with the iterative method introduced by Ref.~\cite{Katz:1991}, where in each iteration the distance of each particle from the halo center is weighted along the eigenvectors with the corresponding eigenvalues of the previous iteration until convergence.
In total, we have used three different halo shape definitions for the Rockstar halos and two different definitions for the AHF halos. We present our main results in reduced shape of the Rockstar halos without iterations, unless otherwise mentioned.

We then construct the shape field from each shape tensor as 
\begin{align}
    S_{ij}({\bf x}) = \frac{I_{ij}({\bf x}) - \langle I_{ij}\rangle}{{\rm Tr}\langle I_{ij}\rangle},
\end{align}
where $I_{ij}({\bf x}) = \sum_\alpha I_{ij}({\bf x}_\alpha)\delta_{\rm D}^{(3)}({\bf x}-{\bf x}_\alpha)$ with ${\bf x}_\alpha$ being the $\alpha$-th halo position.
We compute this shape field on a $512^3$ grid with the standard CIC distance weight.
Note that we normalize the shape field by the \emph{average three-dimensional trace} of the inertia tensor field, rather than the trace of each individual halo or two-dimensional ones after projection.
As a result, the trace of $S_{ij}({\bf x})$ has a fluctuation depending on the position:
\begin{align}
    S_{ij}({\bf x}) =  \frac{1}{3}\delta_{ij}^{\rm K}\delta_s({\bf x}) + g_{ij}({\bf x}),
\end{align}
where $\delta_s$ is the size fluctuation and $g_{ij}$ is the trace-free part of $S_{ij}$ and corresponds to the usual shape field with the other normalization.
In this paper, we consider only the $\ell = 2$ component of the spherical tensor to focus on the trace-free part of $S_{ij}$.

\section{Result}
\label{sec:result}

\begin{figure}[tb]
\centering
\includegraphics[width=1.0\textwidth]{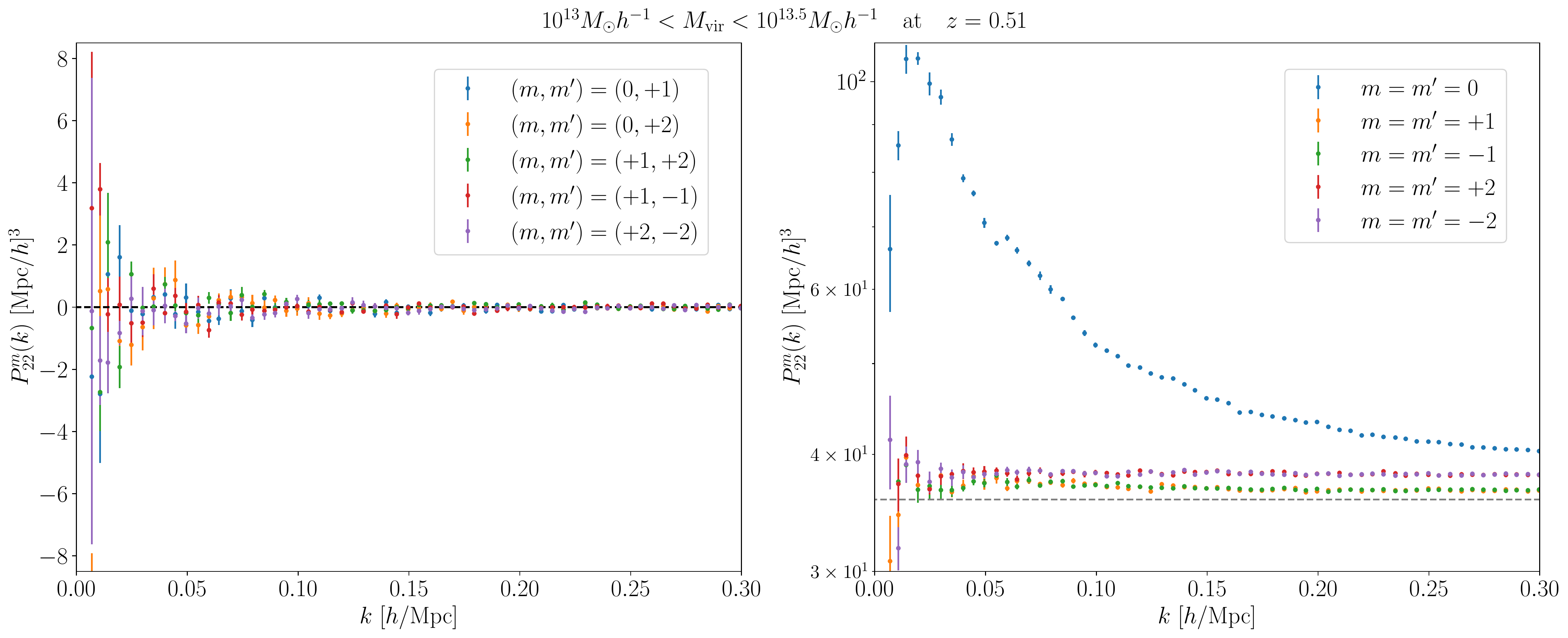}
\hfill
\caption{Shape auto-power spectra in the spherical tensor basis with the halo mass in range $10^{13} M_\odot h^{-1} < M_{\rm vir}  < 10^{13.5} M_\odot h^{-1}$ at $z=0.51$.
The left and right panels show the power spectra for $m \neq m'$ and $m=m'$ cases, respectively.
The black dashed line corresponds to the zero and the grey dashed line corresponds to the shape noise term, $2\sigma^2_\gamma/\Bar{n}_{\rm h}$.
Note that this shape noise is not subtracted in all cases.
}
\label{fig:auto_powers}
\end{figure}

\begin{figure}[tb]
\centering
\includegraphics[width=1.0\textwidth]{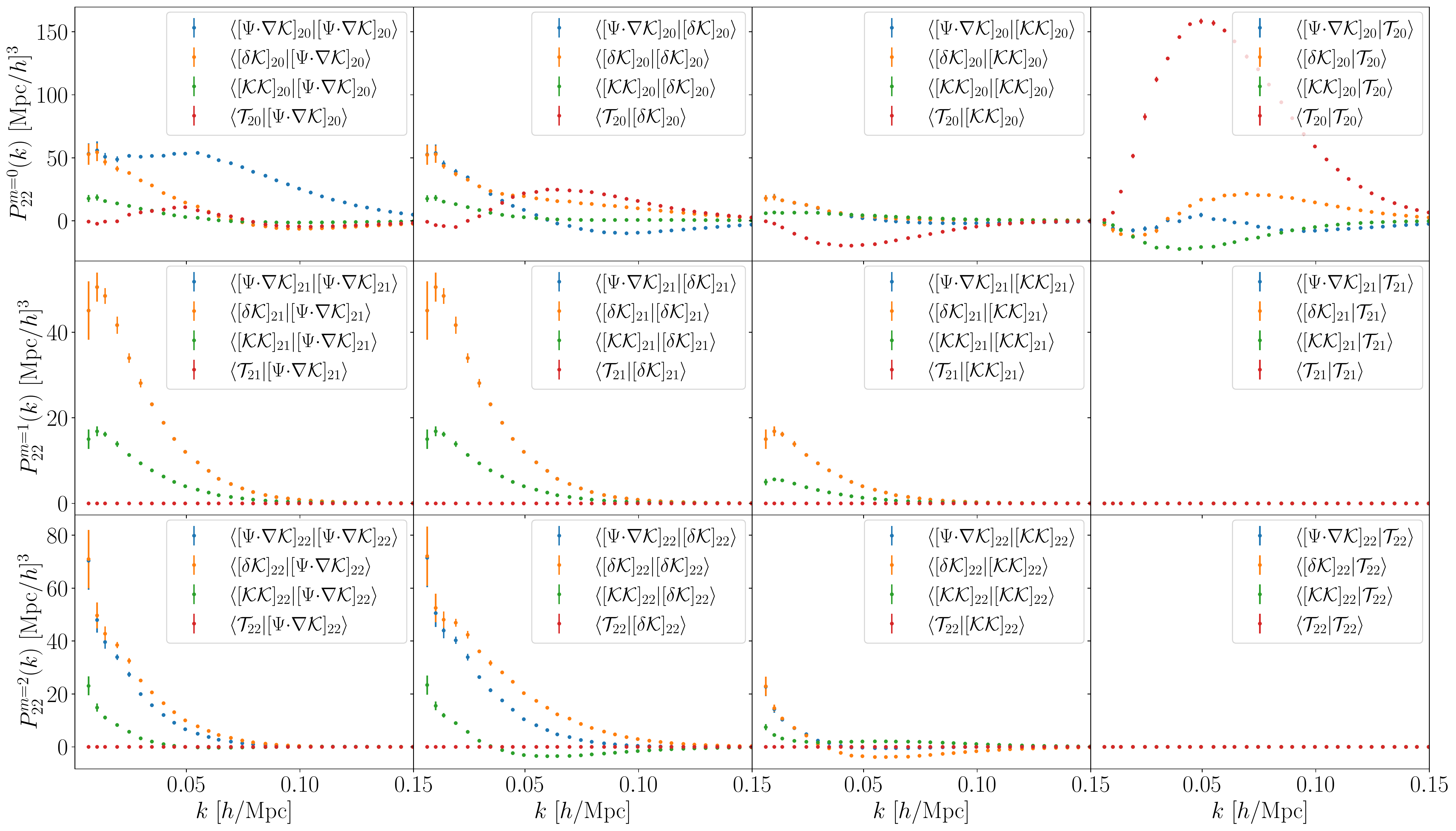}
\includegraphics[width=1.0\textwidth]{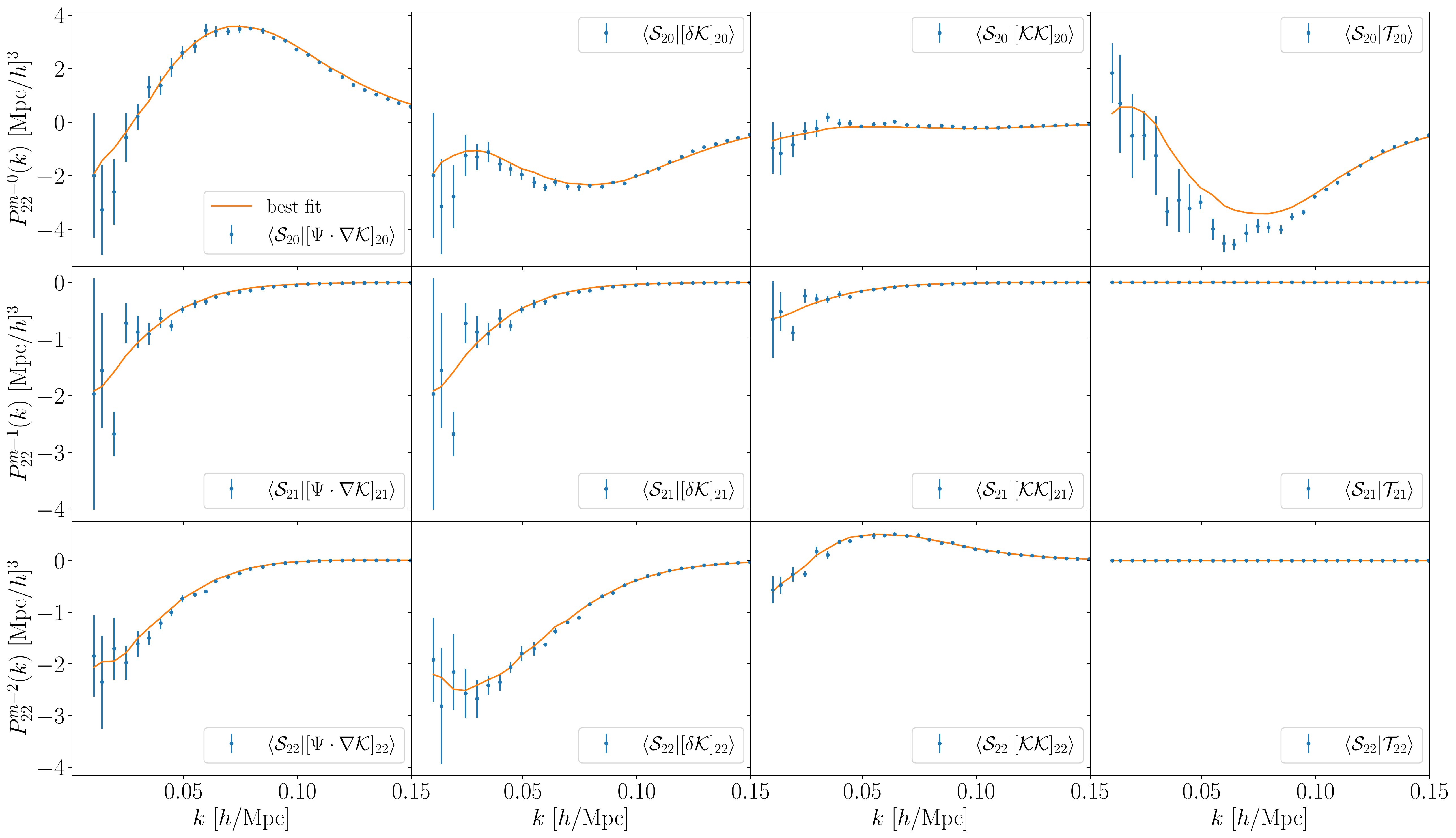}

\caption{
Auto-power spectra of the quadratic fields (top panels) and 
cross-power spectra between the halo shape field and the quadratic fields (bottom panels) with the halo mass in $10^{13} M_\odot h^{-1} < M_{\rm vir}  < 10^{13.5} M_\odot h^{-1}$ at $z=0.51$.
The upper, middle, and lower rows show $P^{m=0}_{22}(k)$, $P^{m=1}_{22}(k)$, and $P^{m=2}_{22}(k)$, respectively.
}
\label{fig:cross_powers}
\end{figure}

In this section, first we present the power spectra of halo shape field and quadratic fields that are decomposed in the spherical tensor basis. Then we show the measurement of the quadratic shape biases using these power spectra.

\subsection{Power spectra in the spherical tensor basis}

Before discussing about the quadratic shape biases, we provide the measurements of the power spectra of the spherical tensors.

Fig.~\ref{fig:auto_powers} shows the auto-power spectra of the halo shape field decomposed by the spherical tensor basis for $10^{13} M_\odot h^{-1} < M_{\rm vir}  < 10^{13.5} M_\odot h^{-1}$ halos at $z=0.51$.
On the left panel, we show the power spectra $\langle {\mathcal S}_{2 m} ({\bf k}) {\mathcal S}_{2 m'} ({\bf k}') \rangle^\prime$ with $m\neq m'$ and on the right we show $m = m'$.
Note that the Poisson noise expectation, $2\sigma^2_\gamma/\Bar{n}_{\rm h}$, is not subtracted in either panel.
First, for $m\neq m'$ the measured power spectra are consistent with zero as expected, which verifies that our code works correctly.
Second, it is also confirmed that $P^m_{22} = P^{-m}_{22}$ for $m=1,2$ since our simulation does not violate the parity symmetry.
Thus the only independent power spectra are $P_{22}^{m=0}$, $P_{22}^{m=\pm 1}$, and $P_{22}^{m=\pm 2}$.
Clearly, $P_{22}^{m=0}$ has largest amplitude because it contains the tree-level (linear) contribution while the others do not.
Going to the smaller scales, however, three power spectra have comparable amplitude even after subtracting the shape noise contribution, implying that into the non-linear correction is important for the intrinsic alignment.

In addition, we find that the power spectra with $m=\pm 1$ and $m=\pm 2$ have non-negligible excess above the Poisson noise expectation (shown by the dashed grey line) even at relatively large scales, meaning that there are indeed non-linear contributions.
We note that up to the one-loop level they come from only the ``22'' contribution (the correlations between the quadratic fields), not the ``13'' contribution (the correlations between the linear and cubic fields), since the linear field ($K_{ij} \propto [Y^{m=0}_{2}(\hat{k})]_{ij}$) involved in the ``13'' contribution is orthogonal to the $m\neq 0$ basis.
Further, these nonlinear contributions should stem from the nonlinear biasing, not the nonlinear evolution of the matter field, because in the linear bias with nonlinear tidal field case (i.e., the nonlinear alignment model~\cite{Bridle:2007ft}) we have $S_{ij}=b_K K_{ij}^{\rm NL}\propto [Y^{m=0}_{2}(\hat{k})]_{ij}$, which is orthogonal to the $m\neq 0$ basis as well.
Hence, the detection of $P_{22}^{m=\pm 1}$ and $P_{22}^{m=\pm 2}$ serves as the clear evidence of the quadratic shape biases and demonstrates that the widely-used nonlinear alignment model misses some nonlinear contributions.

In Fig.~\ref{fig:cross_powers} we present the power spectra among the quadratic fields on the top panel and the cross-power spectra between the halo shape field and the quadratic fields on the bottom panel.
We subtract the ``12'' contributions as described in the previous section when plotting the cross-correlations between the halo shape and the quadratic fields; otherwise they add a significant scatter.
One can see that in the $m=1$ and $m=2$ cases the power spectra involving ${\cal T}$ field vanish since ${\cal T}_{2\pm 1}={\cal T}_{2\pm 2}=0$ as discussed in Sec.~\ref{subsec:ST_decomposition}.
In addition, in the $m=1$ case all the power spectra have essentially the same shape and in particular the power spectra involving $\left[\delta{\mathcal K} \right]_{2 \pm 1} $ and $\left[ \Psi\cdot\nabla{\mathcal K}\right]_{2 \pm 1}$ fields are equivalent, which is expected from Eq.~\eqref{eq:identity_pm1}.
Note that the smoothing with $R=20~{\rm Mpc}/h$ suppresses all the power spectra in the high-$k$ regime.
We estimate the shape biases by fitting the cross-power spectra of halo shape and the quadratic fields with combinations of the quadratic field power spectra each with their own characteristic $k$-dependence.

\subsection{Quadratic shape biases}
\label{subsec:quad_shape_bias}

\begin{figure}[tb]
\centering
\includegraphics[width=1.0\textwidth]{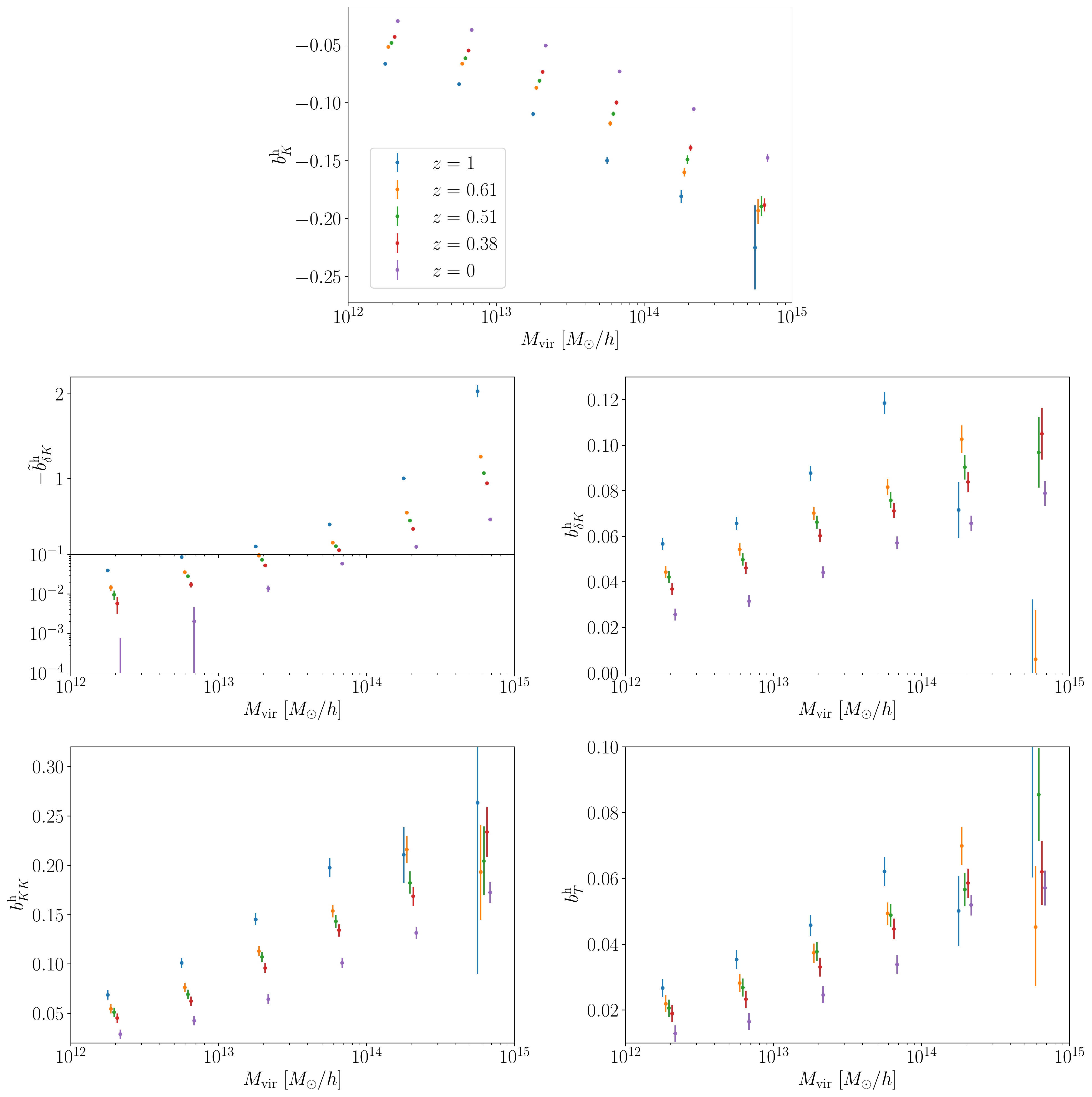}

\caption{
The shape bias parameters as a function of halo mass for various redshifts.
The top panel describes the linear shape bias $b_K$, while the remaining four panels show quadratic shape biases, $-\tilde{b}_{\delta K}$, $b_{\delta K}$, $b_{KK}$, and $b_{T}$, defined in Eq.~\eqref{eq:bias_expansion_Eulerian} and Eq.~\eqref{eq:tilde_bdK}.
The middle left panel shows the shape bias to the density-weighting term ($\delta K_{ij}$) for the density-weighted shape field, $\tilde{b}_{\delta K}$, whereas the middle right panel shows that for the volume-weighted shape field, $b_{\delta K}$.
Note that because the actual values of $\tilde{b}_{\delta K}$ are negative we show $-\tilde{b}_{\delta K}$ in the middle left panel so that 
we can make the logarithmic plot in the small $|\tilde{b}_{\delta K}|$ regime.
}
\label{fig:quad_bias_mass_redshift}
\end{figure}

Here we discuss the measurement of the quadratic shape biases.
One thing to keep in mind is the density weighting discussed in Sec.~\ref{subsec:shape_bias}.
Although we can formally define the shape bias for the volume-weighted shape field as in Sec.~\ref{subsec:shape_bias}, the natural observable is density-weighted shape field.
Thus for the shape bias for the density-weighting term ($\delta K_{ij}$) what we can directly measure is $\tilde{b}_{\delta K} = b_{\delta K} + b_1 b_K$.
On the other hand, in simulations we can precisely measure $b_1$ and $b_K$ of the same halo sample so that we can estimate $b_{\delta K}$ as well.
In the following we show the measurements of both $\tilde{b}_{\delta K}$ and $b_{\delta K}$.

Fig.~\ref{fig:quad_bias_mass_redshift} summarizes the mass- and redshift-dependence of each shape bias parameters.
The linear shape bias, $b_K$, shown in the top panel, is qualitatively consistent with the result in Ref.~\cite{Akitsu:2020fpg}, although the method to measure the linear bias is completely different from this paper.
The middle panels show both $\Tilde{b}_{\delta K}$ and $b_{\delta K}$, and the bottom left and right panels shows $b_{KK}$ and $b_{T}$, respectively.
Clearly we detect all three quadratic shape biases with high significance for all redshifts and halo masses.
Despite the use of only eight realizations, we obtain the small error bars, which is mainly due to the cosmic variance cancellation.
Also we note that these errors correspond to the standard error of the mean of the shape bias parameters, not the standard deviation, as discussed in Sec.~\ref{subsec:quad_field_method}.

The dependence of the quadratic shape biases on redshift and halo mass is similar to that of the linear shape bias; the absolute value of biases is larger for massive halos and at earlier redshift.
The advection argument described in Sec.~\ref{subsec:shape_bias} does indeed anticipate this trend, as in Eq.~\eqref{eq:quad_shape_advection}.
In the massive end at $z=1$ the measurements look off from this trend in the $b_{\delta K}$ and $b_T$ plots, although further investigation would be necessary as the number of halos might not be sufficient to get the accurate result in this regime.
One may notice that the value of $\Tilde{b}_{\delta K}$ is order-of-magnitude larger than the other quadratic biases.
This is because $\Tilde{b}_{\delta K}$ contains $b_1 b_K$ and the density bias $b_1$ is much larger than the shape bias as $b_1$ is ${\cal O}(1)$-${\cal O}(10)$ whereas $b_K$ and other quadratic shape biases are just ${\cal O}(0.1)$.
In fact, if the quadratic shape biases are generated due to the advection discussed in Sec.~\ref{subsec:shape_bias}, they should have comparable amplitude to the linear shape bias $b_K$.

\begin{figure}[tb]
\centering
\includegraphics[width=1.0\textwidth]{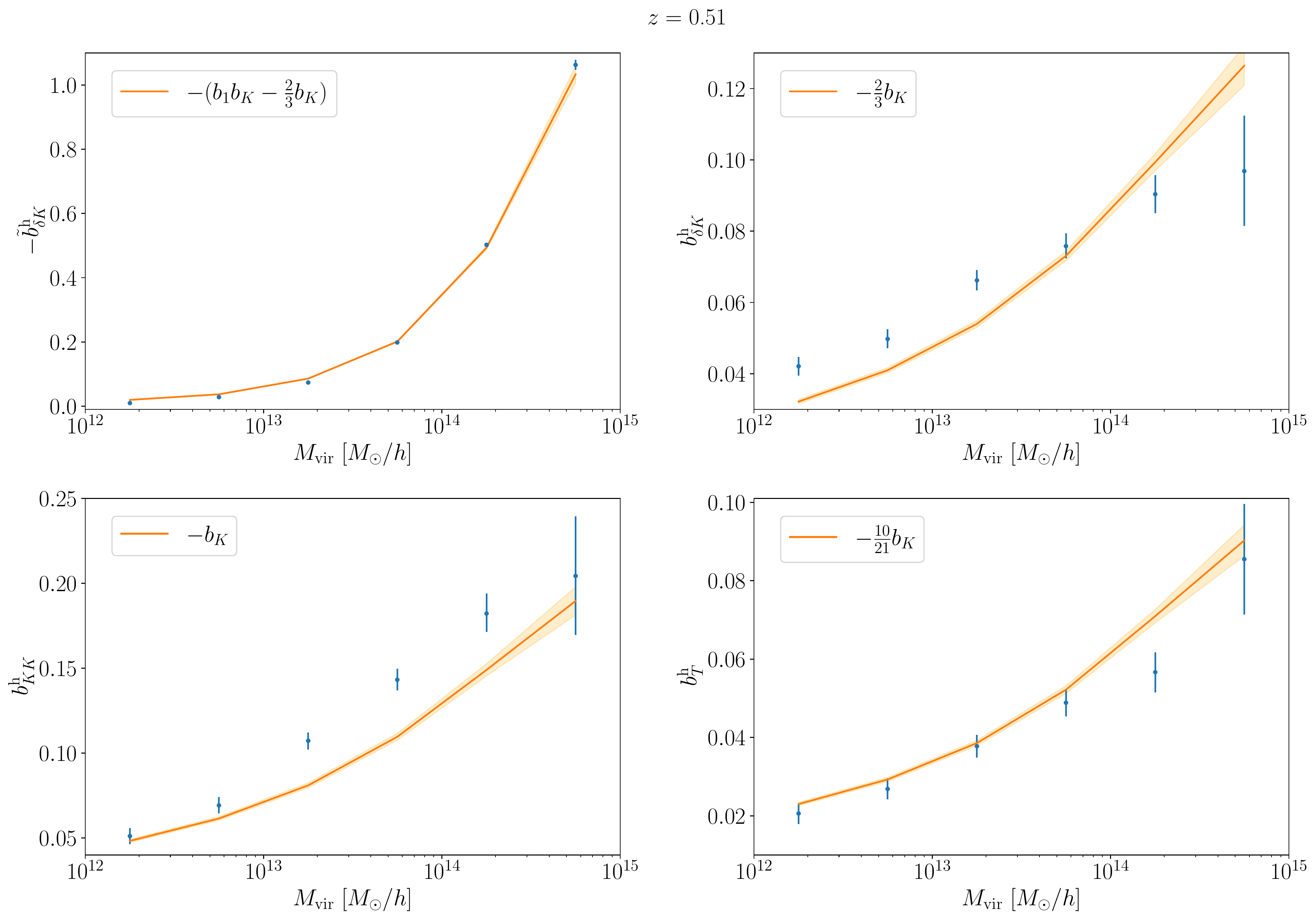}

\caption{
The quadratic shape bias parameters as a function of halo mass at $z=0.51$, with the advection predictions.
The blue points show the measurements from the simulations (the same as Fig.~\ref{fig:quad_bias_mass_redshift}) and the orange line corresponds to the advection prediction (Eq.~\eqref{eq:quad_shape_advection} and Eq.~\eqref{eq:quad_shape_advection_density_weighting}).
The orange shaded region corresponds to the error of the measurement of $b_K$, which is estimated by the same way as the quadratic bias parameters (see Sec.~\ref{subsec:quad_field_method} for the details).
We show $-\tilde{b}_{\delta K}$ in the top left panel, to be consistent with Fig.~\ref{fig:quad_bias_mass_redshift}.
}
\label{fig:quad_bias_bK_masses}
\end{figure}

Let us compare the measurements of the quadratic shape biases with the advection prediction  Eq.~\eqref{eq:quad_shape_advection}.
In Fig.~\ref{fig:quad_bias_bK_masses} we show the measured quadratic shape biases by blue points and the prediction by orange lines as a function of halo mass at $z=0.51$.
As the prediction relates the quadratic shape biases to the linear shape bias $b_K$, we use the measured $b_K$ from the same halo sample to plot the prediction, with the orange bands corresponding to errors on the $b_K$ measurement.
Note that for $\Tilde{b}_{\delta K}$ there appears the additional term $b_1b_K$ in the prediction as we discuss in the beginning of this section (or see Eq.~\eqref{eq:quad_shape_advection_density_weighting}).
The advection prediction provides a reasonably good agreement with the actual measurement.
Looking a little closer, there are small deviations from the prediction for $b_{\delta K}$ and $b_{KK}$, while for $b_T$ it describes the measurements quite well within the errors.
The small departure from the orange lines observed in $b_{\delta K}$ and $b_{KK}$ implies the nonzero Lagrangian quadratic biases.
In other words, at the epoch of the formation the halo shapes are affected by not only the tidal effect but also the density and tidal torquing.

\begin{figure}[tb]
\centering
\includegraphics[width=1.0\textwidth]{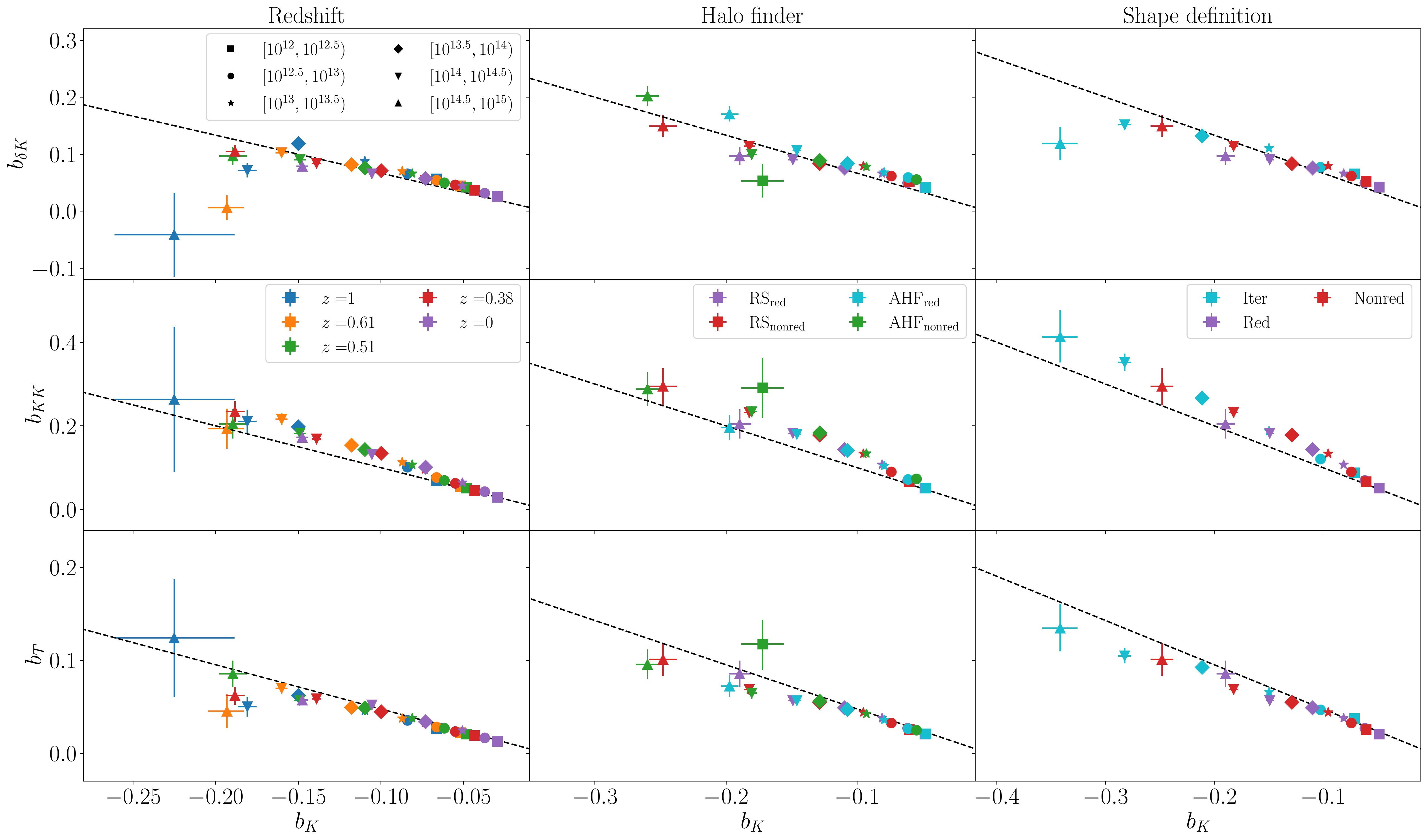}
\caption{
The quadratic shape biases as a function of the linear shape bias for various redshifts (the left), different halo finder (the middle), and various shape definition (the right).
The fiducial choice of these is the halos identified by Rockstar at $z=0.51$ with the shape being the reduced inertia tensor without iterations.
The different symbols correspond to different halo mass.
The dashed lines show the advection prediction for each quadraric shape bias.
}
\label{fig:quad_bias_bK}
\end{figure}

Given these good agreements, it is natural to ask how general it is.
To this end, we investigate the universality of this prediction against redshift, halo finder, and shape definitions.
In Fig.~\ref{fig:quad_bias_bK} we show the quadratic shape biases as a function of $b_K$ with the advection prediction (shown by the dashed black lines).
Different symbols indicate different halo masses, and in the left, the middle, and the right panels we test the universality against redshifts, halo finders, and shape definitions, respectively.
For these comparisons we employ the fiducial choice that is the Rockstar halos at $z=0.51$ with the shape definition being the reduced inertia tensor without iterations.

First, we find that the advection prediction works well across all these variations, meaning that the assumption of the linear alignment in Lagrangian space provides a reasonable approximation no matter what choices are made in the measurements of the intrinsic alignments.
While it is known that a different choice in the measurement of the intrinsic alignments can lead to a different amplitude of the linear shape bias, $b_K$, the choice also affects the quadratic shape biases consistently, so that the relations between the linear and quadratic shape biases remain unchanged.
In particular, for different redshifts (the left panels), the scatter of the measurements is so small that there seems to be a universal relationship independent of redshift, except for the massive halos at high redshifts, which can be attributed to the very small number of halos in this range in our simulations, and should be further investigated.
On the other hand, the scatter from the use of the different halo finder is greater, in particular when using the non-reduced inertia tensor as the shape proxy at the lowest mass bins ($M_{\rm h}=[10^{12}, 10^{12.5})$ shown by the square).
Given that at these lowest mass bins the number of particles in the halo is only hundreds or less, and that AHF uses only spherical overdensity information to determine the member particles while Rockstar uses kinematic information to determine the member particles, the non-reduced inertia tensor of the AHF halos may not be a good proxy for the halo shape.
Excluding these outliers, the scatter arising from the halo finder is modest.
The scatter from different shape definitions is also not that large, although the massive halos look like outliers with the iterative shape measurement, especially in the $b_K$-$b_{\delta K}$ panel. Since the massive halos are likely to be formed through mergers, it is possible that the iterative shape measurement somehow gives some weight to substructures in the massive halos, which could gives rise to a different response to the quadratic bias bases, although this anomaly should also be investigated further.

Second, however, these universal relations observed under different choices have small discrepancy from the advection prediction.
The largest deviation is seen in the $b_K$-$b_{KK}$ relation; in all cases, the measurements of $b_{KK}$ tend to be larger than the prediction, especially for larger $|b_K|$, confirming that the feature seen in Fig.~\ref{fig:quad_bias_bK_masses} is stable across  different choices.
This deviation serves as indirect evidence for the presence of the positive Lagrangian tidal torque bias $b_{KK}^{\rm (L)}$ regardless of different choices.

The deviations are also observed for the other two quadratic biases although they are a bit smaller.
The measurements of $b_{\delta K}$ tend to be higher than the prediction at small $|b_K|$ regime but lower at large $|b_K|$ regime, which again confirms the trend in Fig.~\ref{fig:quad_bias_bK_masses}.
This also indicates the presence of the Lagrangian density-weighting bias $b_{\delta K}^{\rm (L)}$.
It should be noted that there could be another contribution for this term if one uses the conventional shape tensor, i.e., the shape tensor normalized by the \textit{each} trace of the inertia tensor field, instead of the \textit{averaged} trace as we do in this paper.
For the shape tensor normalized by the each trace, the size (trace) fluctuation can enter the shape bias expansion through the normalization, in particular to the density-weighting term at second order.
In other words, the measurement of $b_{\delta K}$ can involve the $b_{s} b_K$ contribution as in the $b_1 b_K$ contribution with $b_s$ being the linear size bias defined via $\delta_{s} = b_{s}\delta$.
Hence the advection prediction would not be accurate for the usual shape tensor unless we take into account the size fluctuation effect with unknown $b_{s}$, although we expect $b_s$ would be small.

Finally, $b_T$ exhibits the small departure from the advection prediction for large $|b_K|$ as well.
The measurements for massive halos tend to get smaller values than the prediction, implying the presence of the negative $b_{T}^{{\rm (L)}}$ for massive halos.

\section{Discussion}
\label{sec:discussion}

For the first time, we present the precise measurements of the quadratic shape bias parameters appeared in the shape bias expansion of the intrinsic alignments.
We utilize the spherical tensor decomposition of the shape field to efficiently exploit the full three-dimensional information on the shape field, instead of projecting onto the plane perpendicular to the line-of-sight as is usually done where some information is lost.
We have measured the power spectra of the shape field in the spherical tensor basis for the first time and confirmed that non-linear biasing produces non-vanishing vector and tensor modes, as in the usual cosmological perturbation theory.
In this basis, we take cross-correlations of the quadratic shape bias operators from the Gaussian initial conditions with the halo shape field that shares the random seeds.
These two methods we employ allows for the precise measurements of the quadratic shape biases.

We find clear evidence for nonzero quadratic shape biases in Eulerian space. 
We also compare the amplitude of the detected quadratic shape biases with the prediction where only the linear shape bias is present in Lagrangian space and higher-order shape biases in Eulerian space are dynamically generated via the displacement.
Our measurements are generally consistent with this prediction, although for all three quadratic biases we have detected small deviations.
The small but non-negligible deviations in turn implies the existence of nonzero Lagrangian quadratic shape biases, indicating that the halo shape is determined by not only the linear tidal field but also tidal torquing and density field at the formation epoch.
We have also validated that this trend is stable against different redshifts, different halo masses, different halo finders, and different shape definitions.

The general agreements between the measurements of the quadratic shape biases and the advection prediction suggest that we can put the priors on the quadratic shape biases from the advection prediction.
This would be valuable for the perturbative modeling of the intrinsic alignments, whether the intrinsic alignments are treated as a cosmological signal or as a contamination to the weak lensing.
While there are small deviations from the advection prediction, these deviations would not matter for actual applications, given the small signal-to-noise of the intrinsic alignments and necessity to marginalizing over the unknown initial conditions, which we assume known to make the precise measurements in this work.

The methodology presented here can be straightforwardly extended to the measurements of the cubic shape biases, which complete the one-loop power spectrum.
To this end, one can consider the cross-correlation between the cubic fields with the halo shape field, or the bispectrum of the linear field, the quadratic fields, and halo shape fields.
We leave this study for future work.
In addition, the use of spectra of the spherical tensor also provides an efficient stress test for the model of the intrinsic alignments because in this basis we do not lose any information about the shape field, whereas the projected spectra do not contain full information of the intrinsic alignment.
Although the actual observables are projected quantities, the intrinsic alignments itself is three-dimensional phenomenon and hence the full three-dimensional information would be more valuable for studying the intrinsic alignments in simulations.

\acknowledgments
The authors thank Shi-Fan Chen for useful discussions.
KA is supported by JSPS Overseas Research Fellowships.
TO acknowledges support from the Ministry of Science and Technology of Taiwan under Grants No. MOST 111-2112-M-001-061- and the Career Development Award, Academia Sinica (AS-CDA-108-M02) for the period of 2019-2023. 
Numerical computation was carried out on the Helios and the Typhon cluster at the Institute for Advanced Study and the Cray XC50 at the Center for Computational Astrophysics, National Astronomical Observatory of Japan.

\appendix

\section{Derivation of Eq.~\eqref{eq:identity_pm1}}
\label{app:proof}

In order to prove Eq.~\eqref{eq:identity_pm1}, we first consider the following identity,
\begin{align}
  {\bf \Psi}\cdot\nabla K_{ij} & = - \left(\nabla\cdot{\bf \Psi}\right)K_{ij} + \nabla\cdot\left({\bf \Psi} K_{ij}\right)
    \nonumber
    \\
    & = \delta K_{ij} +  \nabla\cdot\left({\bf \Psi} K_{ij}\right).
\end{align}
Thus, the difference between ${\bf \Psi}\cdot\nabla K_{ij}$ and $\delta K_{ij}$ is given by $\nabla\cdot\left({\bf \Psi} K_{ij}\right)$.
In Fourier space this term is expressed as 
\begin{align}
    \left[ \nabla\cdot\left({\bf \Psi} K_{ij} \right) \right]({\bf k})
    = \int_{\bf q} ~\frac12\left[\frac{{\bf k}\cdot{\bf p}}{p^2}\left(\frac{q_i q_j}{q^2} - \frac13 \delta_{ij}^{\rm K}\right)
    + \frac{{\bf k}\cdot{\bf q}}{q^2}\left(\frac{p_i p_j}{p^2} - \frac13 \delta_{ij}^{\rm K}\right)
    \right]\delta({\bf p})\delta({\bf q}),
\end{align}
with ${\bf p} = {\bf k} - {\bf q}$.
Focusing on the inside of the bracket in the above and projecting this term onto $[Y_{\ell=2}^{m=\pm 1}(\hat{k})]_{ij}$ basis yields 
\begin{align}
    \left[\frac{{\bf k}\cdot{\bf p}}{p^2}\left(\frac{q_i q_j}{q^2} - \frac13 \delta_{ij}^{\rm K}\right)
    + \frac{{\bf k}\cdot{\bf q}}{q^2}\left(\frac{p_i p_j}{p^2} - \frac13 \delta_{ij}^{\rm K}\right)
    \right]
    \left[Y_{\ell=2}^{m=\pm 1}({\hat{k}})\right]_{ij} 
    = \sqrt{2} \frac{k\mu_p\mu_q}{pq}\left({\bf p}+{\bf q}\right)\cdot{\bf e}^\pm
    =0,
\end{align}
where we have defined $\mu_p = \hat{k}\cdot\hat{p}$ and $\mu_q = \hat{k}\cdot\hat{q}$.
Hence we have 
\begin{align}
    \left[\delta{\mathcal K} \right]_{2 \pm 1} 
    = \left[ \Psi\cdot\nabla{\mathcal K}\right]_{2 \pm 1}.
\end{align}
Combining this result with Eq.~\eqref{eq:consistency_ST_basis} leads to 
\begin{align}
    \left[{\mathcal K{\mathcal K}} \right]_{2\pm 1}
    = \frac13 \left[\delta{\mathcal K} \right]_{2 \pm 1} 
    = \frac13 \left[ \Psi\cdot\nabla{\mathcal K}\right]_{2 \pm 1}.
\end{align}




\bibliographystyle{JHEP}
\bibliography{main}

\end{document}